# TOOLS AND TECHNIQUES FOR NETWORK FORENSICS

Natarajan Meghanathan, Sumanth Reddy Allam and Loretta A. Moore

Department of Computer Science, Jackson State University, Jackson, MS 39217, USA
[1]nmeghanathan@jsums.edu, [2]sumanth.project@gmail.com,
[3]loretta.a.moore@jsums.edu

## ABSTRACT

*Network forensics deals with the capture, recording and analysis of network events in order to discover evidential information about the source of security attacks in a court of law. This paper discusses the different tools and techniques available to conduct network forensics. Some of the tools discussed include: eMailTrackerPro – to identify the physical location of an email sender; Web Historian – to find the duration of each visit and the files uploaded and downloaded from the visited website; packet sniffers like Ethereal – to capture and analyze the data exchanged among the different computers in the network. The second half of the paper presents a survey of different IP traceback techniques like packet marking that help a forensic investigator to identify the true sources of the attacking IP packets. We also discuss the use of Honeypots and Honeynets that gather intelligence about the enemy and the tools and tactics of network intruders.*

## 1. INTRODUCTION

Internet usage has increased drastically in the past ten years. Recent studies reveal that today in the United States for every three people, one would be using the Internet for their personal activity, or for their business needs [1]. As the number of people using the Internet increases, the number of illegal activities such as data theft, identity theft, etc also increases exponentially. Computer Forensics deals with the collection and analysis of data from computer systems, networks, communication streams (wired and wireless) and storage media in a manner admissible in a court of law [2]. Network forensics deals with the capture, recording or analysis of network events in order to discover evidential information about the source of security attacks in a court of law [3]. With the rapid growth and use of Internet, network forensics has become an integral part of computer forensics. This paper surveys the tools and techniques (efficient, easy to use and cost effective) available to conduct network forensics.

Section 2 explains how to conduct "Email Forensics" using certain freely available tools such as *EmailTrackerPro* [4] and *SmartWhoIs* [5]. Spam emails are a major source of concern within the Internet community. The tools described in this Section could be used to trace the sender of an email. Section 3 describes how to conduct "Web Forensics" using freely available tools like *Web Historian* [6] and *Index.dat analyzer* [7]. These tools help to reveal the browsing history of a person including the number of times a website has been visited in the past and the duration of each visit, the files that have been uploaded and downloaded from the visited website, the cookies setup as part of the visits and other critical information. Section 4 describes the use of packet sniffers like *Ethereal* [8] to explore the hidden information in the different headers of the TCP/IP protocol stack. These sniffers capture the packets exchanged in the Ethernet and allow the investigator to collect critical information from the packets.

The second half of the paper presents a survey on the different techniques proposed for network forensics in research articles. Section 5 describes the use of IP traceback techniques to reliably determine the origin of a packet in the Internet, i.e., to help the forensic investigator to identify the true sources of the attacking IP packets [9]. The IP traceback techniques allow a victim to identify the network paths traversed by the attack traffic without requiring interactive operational support from Internet Service Providers (ISPs). Section 6 describes the use of Honeypots and Honeynets in intrusion detection systems [10, 11]. A Honeypot is a computer in a network to which no traffic should come from outside. A network of Honeypots across the Internet is called a Honeynet. As no traffic from outside should arrive to a Honeypot, if one notices packets arriving to a Honeypot, it is an indication of an attack being launched on the network in which the Honeypot is located. This information could be used to counter threats or an imminent attack. Section 7 presents the conclusions.

## 2. EMAIL FORENSICS

Email is one of the most common ways people communicate, ranging from internal meeting requests, to distribution of documents and general conversation. Emails are now being used for all sorts of communication including providing confidentiality, authentication, non-repudiation and data integrity. As email usage increases, attackers and hackers began to use emails for malicious activities. Spam emails are a major source of concern within the Internet community. Emails are more vulnerable to be intercepted and might be used by hackers to learn of secret communication. Email forensics refers to studying the source and content of electronic mail as evidence, identifying the actual sender and recipient of a message, date/time it was sent and etc. Emails frequently contain malicious viruses, threats and scams that can result in the loss of data, confidential information and even identity theft. The tools described in this section provide an easy-to-use browser format, automated reporting and easy tool bar access features. The tools help to identify the point of origin of the message, trace the path traversed by the message (used to identify the spammers) and also to identify the phishing emails that try to obtain confidential information from the receiver.

*eMailTrackerPro* [4] analyzes the header of an email to detect the IP address of the machine that sent the message so that the sender can be tracked down. All email messages contain a header, located at the top of the email. The header contains the source of an email in the "From" line, while in the "Received" lines, the header lists every point the email passed through on its journey, along with the date and time. The message header provides an audit trail of every machine the email has passed through. The built-in location database in *eMailTrackerPro* helps to track emails to a country or region of the world, showing information on a global map. To trace an email message, one has to just copy and paste the header of the email in *eMailTrackerPro* and start the tool. A basic trace will be shown on the main Graphical User Interface and a summary report can be obtained. The summary report provides an option to report the abuse of the particular email address to the administrators of the sender and/or victim networks and also contains some critical information that can be useful for forensic analysis and investigation. The report includes the geographic location of the IP address from which the email was sent, and if this cannot be found, the report at least includes the location of the target's ISP. The report also includes the domain contact information of the network owner or the ISP, depending on the sender email address. The domain registration details provide information such as who has registered the website address, from what time and how many emails have been sent from that address, and etc.

*SmartWhoIs* [5] is a freeware network utility to look up all the available information about an IP address, hostname or domain, including country, state or province, city, name of the network provider, administrator and technical support contact information. Sometimes, one has to use *SmartWhoIs* along with *eMailTrackerPro* if the information provided by the latter is not complete. Once installed on a machine, a *SmartWhoIs* button will be added to the Internet Explorer (IE) toolbar. Using this handy button, one can invoke *SmartWhoIs* anytime one wants to get more information about the website currently being visited. Using *SmartWhoIs*, one can also query about multiple IP addresses, hostnames or domains at a time.

## 3. WEB FORENSICS

The predominant web browsers in use today are Microsoft's Internet Explorer (IE) and the Firefox/ Mozilla/ Netscape family. Each of these browsers saves, in their own unique formats, the web browsing activity (also known as web browsing history) of the different users who have accounts on a machine. IE stores the browsing history of a user in the index.dat file and the Firefox/ Mozilla/ Netscape family browsers save the web activity in a file named history.dat. These two files are hidden files. So, in order to view them, the browser should be setup to show both hidden files and system files. One cannot easily delete these two files in any regular way. There is also no proof that deleting these files has sped up the browsing experience of the users. Web forensics deals with gathering critical information related to a crime by exploring the browsing history of a person, the number of times a website has been visited, the duration of each visit, the files that have been uploaded and downloaded from the visited website, the cookies setup as part of the visit and other critical information.

Mandiant *Web Historian* [6] assists users in reviewing web site URLs that are stored in the history files of the most commonly used web browsers. The tool allows the forensic examiner to determine what, when, where, and how the intruders looked into the different sites. *Web Historian* can be used to parse a specific history file or recursively search through a given folder or drive and find all the browser history files that the tool knows how to parse. *Web Historian* generates a single report (that can be saved in different file formats) containing the Internet activity from all of the browser history files it is able to locate.

The *Index.dat analyzer* [7] is a forensic tool to view, examine and delete the contents of index.dat files. The tool can be used to simultaneously or individually view the browsing history, the cookies and the cache. The tool provides support to directly visit the website listed in the output of the analyzer and also to open the file uploaded to or downloaded from the website. The tool also provides critical information about a cookie like its key-value pair, the website address associated with the cookie, the date/time the cookie was first created and last accessed and etc. In addition to these two tools, there exist tools like *Total Recall* [12], which can be used to extract the list of favourite websites stored in the browser history.

## 4. PACKET SNIFFERS

A sniffer is software that collects traffic flowing into and out of a computer attached to a network [13]. Network engineers, system administrators and security professionals use sniffers to monitor and collect information about different communications occurring over a network. Sniffers are used as the main source for data collection in Intrusion Detection Systems (IDS) to match packets against a rule-set designed to notify anything malicious or strange. Law enforcement agencies use sniffers to gather specific traffic in a network and use the data for investigative analysis.

### 4.1 Ethereal

*Ethereal* is an open source software and widely used as a network packet analyzer. It captures packets live from the network. It displays the information in the headers of all the protocols used in the transmission of the packets captured. It filters the packets depending on user needs. *Ethereal* allows search for packets with some specifications. It gives better representation to understand the results easily by using a colorized display of packets belonging to different protocols. The platforms supported by *Ethereal* include Microsoft Windows, UNIX and Linux. *Ethereal* provides an option to select the interface across which one wants to capture the packets. *Ethereal* can also be run in promiscuous mode to capture packets that are not addressed to the machine running *Ethereal*. While packet capture is under progress, *Ethereal* shows the kind of packets captured along with their protocols like Transport Control Protocol (TCP), User Datagram Protocol (UDP), Address Resolution Protocol (ARP) and etc. When the capturing process is done, it displays the packet list with all the detailed information (refer Figure 1).

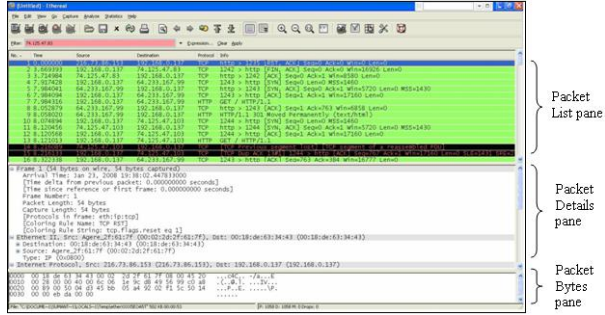

Figure 1. Screenshot of Packet List Displayed by *Ethereal*

The output obtained from *Ethereal* is organized in three panes: Packet List pane, Packet Details pane and Packet Bytes pane. The Packet List pane displays all the packets in the current capture file. Each line in this list corresponds to a packet in the capture file. When one selects a particular line in this list, more details will be listed in the "Packet Details" and "Packet Bytes" panes. The Packet List pane displays high-level details of the packets in several columns. Though, there are a lot of different columns available, the default columns are as follows: (i) No: The number of the packet in the capture file; (ii) Time: The timestamp of the packet; (iii) Source: The IP address of the sender of the packet; (iv) Destination: The IP address of the system to which the packet is destined to go and (v) Protocol: The high-level protocol name in a short abbreviated form.

The Packet Details pane shows the structure of the different headers (starting from the frame-level) of the packet selected in the Packet List pane. Each header can be expanded and collapsed. When one selects a particular header in the Packet Details pane, the byte-level content of the header is displayed in the Packet Bytes pane in a Hexdump style. Generally in Hexdump, the left side shows the offset in the packet data, in the middle the packet data is shown and on the right side, the corresponding ASCII characters are displayed.

*Ethereal* provides options to save the capture file, export the results and also to search for packets based on a specific field or value in the current capture file or a saved capture file. Another useful option provided by *Ethereal* is packet colorization. One can setup the packet colors according to a filter. This helps to easily identify and explore the packets one is usually interested in.

### 4.2 WinPcap and AirPcap

*WinPcap* [14] is the packet capture tool used to capture the packets intercepted at the network interface of a system running the Windows Operating System. *WinPcap* is the tool used for link-layer network access in Windows. *WinPcap* includes a network statistics engine and provides support for kernel-level packet filtering and remote packet capture.

*AirPcap* [15] is the packet capture tool for the IEEE 802.11b/g Wireless LAN interfaces. This tool is currently available only for Windows systems. *AirPcap* can be used to capture the control frames (ACK, RTS, CTS), management frames (Beacon, Probe Requests and Responses, Authentication) and data frames of the 802.11 traffic. The *AirPcap* adapter captures the per-packet power information, which can be used to detect weak signal areas and measure the transmission efficiency of the wireless devices.

## 5. IP TRACEBACK TECHNIQUES

Masquerade attacks [9] can be produced by spoofing at the link-layer (e.g., using a different MAC address than the original), at the Internet layer (e.g., using a different source IP address than the original), at the transport layer (e.g., using a different TCP/IP port than the original one), at the application layer (e.g., using a different email address than the original). Let $C = h_1 \rightarrow h_2 \rightarrow \ldots \rightarrow h_i \rightarrow h_{i+1} \rightarrow \ldots \rightarrow h_n$ be the connection path between hosts $h_1$ to $h_n$. Then, the IP traceback problem is defined as: Given the IP address $h_n$, identify the actual IP addresses of hosts $h_{n-1}, \ldots, h_1$. If $h_1$ is the source and $h_n$ is the victim machine of a security attack, then $C$ is called the attack path [9].

Reconstruction of the attack path back to the originating attacker $h_1$ may not be a straightforward process because of possible spoofing at different layers of the TCP/IP protocol stack and also the intermediate hosts becoming compromised hosts, called stepping-stone, and acting as a conduit for the attacker's communication. The security functions practiced in existing networks may also preclude the capability to follow the reverse path. For example, if the attacker lies behind a firewall, then most of the traceback packets are filtered at the firewall and one may not be able to exactly reach the attacker. The link testing techniques start the traceback from the router closest to the victim and interactively determine the upstream link that was used to carry the attack traffic. The technique is then recursively applied on the upstream routers until the source is reached. Link testing assumes the attack is in progress and cannot be used "post-mortem". There are two varieties of link testing techniques: input debugging and controlled flooding.

### 5.1 Input Debugging

After recognizing that it is being attacked, the victim develops an attack signature that describes a common feature contained in all the attack packets. The victim communicates this attack signature to the upstream router that sends it the attack packets. Based on this signature, the upstream router employs filters that prevent the attack packets from being forwarded through an egress port and determines which ingress port they arrived on. The process is then repeated recursively on the upstream routers, until the originating site is reached or the trace leaves the boundary of the network provider or the Internet Service Provider (ISP). From now on, the upstream ISP has to be contacted to repeat the procedure. The most obvious problem with the input debugging approach, even with the automated tools, is the considerable management overhead at the ISP level to communicate and coordinate the traceback across domains.

### 5.2 Controlled Flooding

The victim uses a pre-generated map of the Internet topology to iteratively select hosts that could be coerced to flood each of the incoming links of the upstream router. Since the router buffer is shared by packets coming across all incoming links, it is possible that the attack packets have a higher probability of being dropped due to this flooding. By observing changes in the rate of packets received from the attacker, the victim infers the link through which the attack packet would have come to the upstream router. This basic testing procedure is then recursively applied on all the upstream routers until the source is reached. Though this method is both ingenious and pragmatic, using unsuspecting hosts to flood is itself a denial-of-service attack. It would be a tremendous overhead to use flooding to detect distributed denial-of-service attacks when multiple upstream links may be contributing to the attack. The victim also needs to possess an accurate topology map to select the hosts that would flood the upstream routers.

### 5.3 ICMP Traceback

In [16], Bellovin proposed the *iTrace* scheme, which is currently one of the IETF standards. According to *iTrace*, every router will sample, with a low probability (like 0.00005), one of the packets it is forwarding and copy the contents into an ICMP traceback message, which will have information about adjacent routers and the message will be sent to the destination. This technique is more likely to be applicable for attacks that originate from a few sources and are of flooding-style attacks so that the receiver gets enough packets to reconstruct the attack path. To handle the situation of attackers sending their own ICMP traceback messages to mislead the destination, the *iTrace* scheme uses HMAC [17] and supports the use of X.509 Digital Certificates [18] for authenticating and evaluating ICMP traceback messages. A couple of potential drawbacks for this scheme are that: (1) ICMP traffic is increasingly getting filtered compared to normal traffic and (2) If certain routers in the attack path are not enabled with *iTrace*, the destination would have to reconstruct several possible attack paths that have a sequence of participating routers, followed by one or more non-participating routers and followed by a sequence of participating routers and so on.

### 5.4 Packet Marking Techniques

The idea behind the packet marking techniques is to sample the path one node at a time rather than recording the entire path. A "node" field, large enough to hold a single router address, in the packet header is reserved. For IPv4, this would be a 32-bit field in the Options portion of the IP header. Upon receiving a packet, a router chooses to write its own address in the node field with a probability *p*. Given that enough packets could be sent and the route remains stable, the victim would receive at least one sample for every router in the attack path.

Since the routers are ordered serially, the probability that a packet will be marked by a router and not marked by the downstream routers is a strictly decreasing function of the distance to the victim. Assuming the probability of marking *p* is the same for every router, the probability of receiving a packet marked from a router *d* hops away and not marked by any other router since then is $p(1-p)^{d-1}$. Figure 2 illustrates the probability of receiving a packet marked from a router 1, 2, 3, 4, 5 and 6 hops away and not marked by any other router on a 6-hop path for different values of the individual probability of marking *p*.

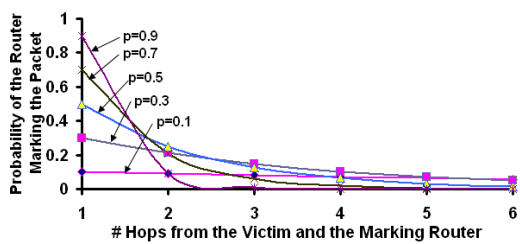 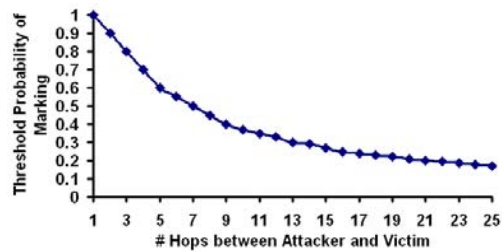

Figure 2. Probability of Marking by a Router   Figure 3. Threshold Probability of Marking
       Vs Hop Count of Attack Path                             Vs Hop Count of Attack Path

The threshold probability of marking is defined as the minimum probability value to be assigned to every router on a path in order to guarantee with 99% probability that at least one router on the path will mark a packet. The threshold probability of marking decreases with the increase in the number of hops. The larger the number of intermediate routers, the greater is the chance of at least one router in the path deciding to mark the packet. Figure 3 shows the threshold probability of marking as the number of hops is varied from 1 to 25.

The convergence time is defined as the minimum threshold number of packets required to determine the sequence of routers that form the attack path. To determine the order of the routers in the attack path, each router on the path should have marked different number of times on the packets. The router that is closest to the victim will have the highest number of marks and the router that is closest to the attacker will have the minimum number of marks. In general, to determine an $n$-hop attack path Attacker → $R_n$ → $R_{n-1}$ → … → $R_i$ → $R_{i-1}$ … → $R_2$ → $R_1$ → Victim, the following two conditions should be satisfied: (i) the victim should receive packets such that every router in the attack path should have marked at least one packet and (ii) the number of packets marked by Router $R_i$ should be strictly greater than $R_{i-1}$, for any $2 \leq i \leq n$. The convergence time is basically the minimum number of packets the victim needs to receive such that the above two conditions are satisfied.

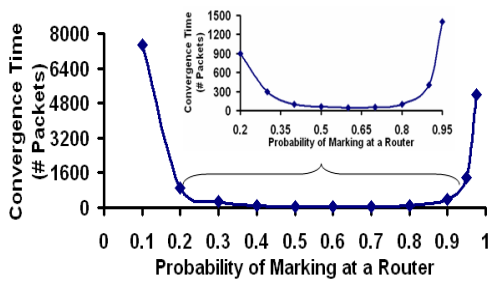

**Figure 4.** Convergence Time for a 3-hop Attack Path

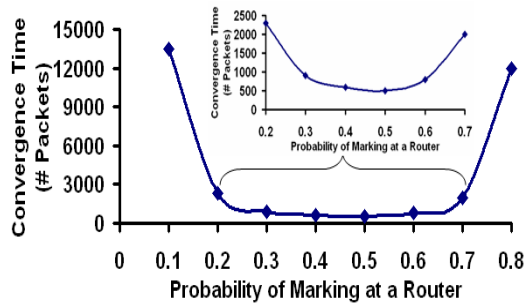

**Figure 5.** Convergence Time for a 6-hop Attack Path

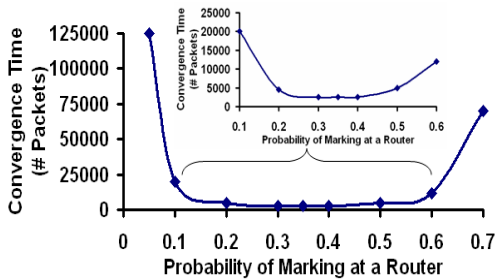

**Figure 6.** Convergence Time for a 9-hop Attack Path

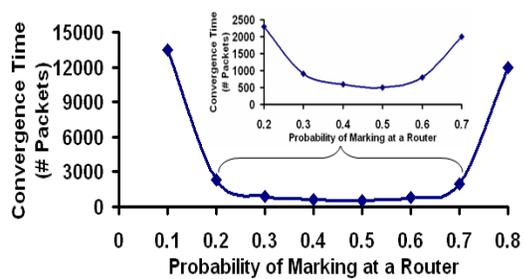

**Figure 7.** Convergence Time for a 12-hop Attack Path

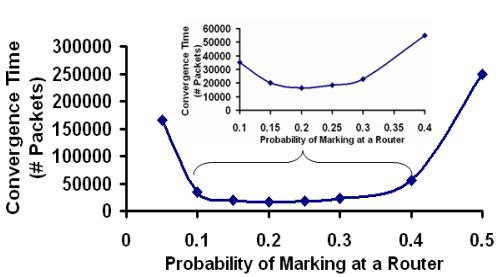

**Figure 8.** Convergence Time for a 15-hop Attack Path

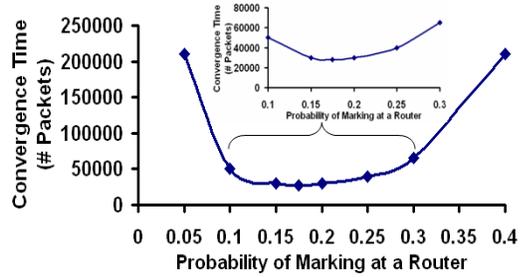

**Figure 9.** Convergence Time for a 18-hop Attack Path

The value of the convergence time depends on the probability of marking by a router and the hop count of the attack path. Figures 4 through 9 show the convergence time measured for different probability of marking values on attacks paths with different hop counts, measured with 95% to 97% confidence intervals. For a given hop count of the attack path, the convergence time is minimum for a certain range of values for the probability of marking. The value of the threshold marking probability decreases as the hop count of the attack path is increased because the probability of any router on the attack path marking the packet increases as the hop count increases. Thus, it is possible to reduce the threshold probability of marking a packet as the hop count increases. The minimum convergence time also increases as the hop count of the attack path increases. This is because as the hop count increases, it takes more time for a router closer to the attacker to have a packet marked such that the packet is not marked by any downstream router on the attack path. In order to lower the convergence time in larger hop count attack paths, it is essential to assign a lower probability of marking for the routers.

**5.5 Source Path Isolation Engine (SPIE) Architecture**

The Source Path Isolation Engine (SPIE) architecture [19] implements the log-based IP traceback technique to trace the path for a single packet across autonomous systems. In SPIE, routers compute a 32-bit digest of the packets based on the 20 byte IP header and the first 8 bytes of the payload. The routers store these 32-bit packet digests, instead of the packets themselves, in a space-efficient data structure called Bloom filter [20].

Within each autonomous system (AS), SPIE consists of three major architectural components: Data Generation Agent (DGA), SPIE collection and reduction agent (SCAR) and SPIE Traceback Manager (STM). The DGA is in charge of computing the 32-bit digests and storing them in the Bloom filter. The SCAR is responsible for generating the attack path within a region of the AS. When requested for an attack path, the SCAR queries the DGAs in the region and compiles the attack path based on the replies from the DGAs. The STM controls the whole SPIE system in the AS. When the STM receives a request to trace an IP packet, it generates requests to the SCARs and then grafts the attack paths returned by the SCARs to form a complete attack path through the AS. As illustrated in Figure 10 and explained below, it is possible to trace an attack path for a single packet that traversed through a mix of SPIE-deployed and non SPIE-deployed ASes.

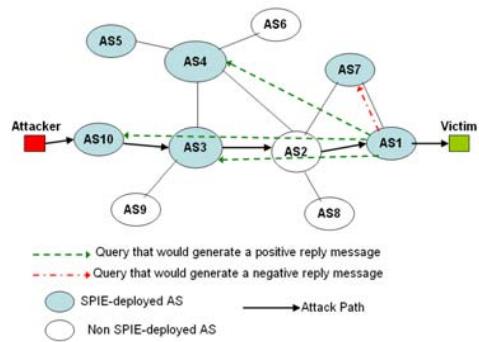

Figure 10. Traceback across Autonomous Systems using SPIE (adapted from [19])

It is assumed that all SPIE-deployed ASes exchange the SPIE deployment information as an attribute in the network route advertisement messages of the Border Gateway Protocol [21]. As a result, each SPIE-deployed AS knows about the other SPIE-deployed ASes, the number of hops (AS levels) and their respective STMs. In Figure 10, an attack packet is sent from an attacker in AS10 to a victim in AS1. When the victim reports about the attack packet to the STM in AS1, the STM queries its one-hop STMs in other ASes about the attack path. Note that AS1 cannot directly contact AS2 as the latter is not SPIE-deployed. So, the STM in AS1 will first query the STM in AS7. Since the attack packet did not traverse through this AS, the STM in AS7, after an internal traceback, sends a negative reply message to the STM in AS1. The STM in AS1 then queries the two-hop STMs in AS3 and AS4. After an internal traceback, these two STMs send back positive reply messages that contain the attack path within their AS. The two ASes will also let AS1 know who are their one-hop and two-hop SPIE-deployed ASes so that AS1 can query the STMs of those ASes. AS3 will let AS1 know that AS10 and AS4 are its one-hop SPIE-deployed ASes and that AS5 and AS7 are two-hop SPIE-deployed ASes. The STM of AS1 contacts its counterpart in AS10. The STM in AS10 will then conduct an internal trace and report back to the STM in AS1 about the attack path that includes the attacker in its AS. The STM of AS1 grafts the attack path in AS1 and the attack paths returned from AS3 and AS10 to get the entire attack path.

The SPIE approach is scalable in the Internet because of the following two reasons: (i) Each STM could be configured with the furthest level of SPIE deployment information to maintain. Generally, STMs need to maintain SPIE deployment information only about nearby ASes (like one-hop, two-hop ASes). A remote STM will notify the victim STM about the nearby STMs to the remote AS and the victim STM can continue querying those STMs as explained before. (ii) An STM need not keep track of the status of ongoing traceback processes.

## 6. HONEYPOTS AND HONEYNETS

A Honeynet is a network specifically designed for the purpose of being compromised [11]. A compromised Honeynet can be used to observe the activities and behaviour of the intruder. A Honeynet allows to conduct a detailed analysis of the tools used by the intruders and to identify the vulnerabilities exploited by the intruders to compromise the Honeypots. In general, a Honeypot could range from a simple port-listening socket program to a full production system that can be emulated under various operating systems. Sometimes, a Honeypot would attract traffic by acting as a decoy system, posing itself to the Internet as a legitimate system offering services. The essential idea is that any traffic directed to the Honeypot is considered an attack or intrusion. Any connection initiated inbound to Honeypot is likely to be a probe, scan or attack. Any outbound connection from a Honeypot implies someone has compromised the Honeypot and has initiated outbound activity. Hence, forensic analysis of the activities of a Honeypot is less likely to lead to false negatives and false positives when compared to the more evasive and signature-dependent network intrusion detection systems [22]. Honeypots can help us to detect vulnerabilities that are not yet understood or not yet seen before.

Honeypots can be classified to two types [23] depending on the configured services available for an adversary to compromise or probe the system: low-interaction and high-interaction Honeypots. A high-interaction Honeypot can be compromised completely, thus allowing an adversary to gain access to all aspects of the operating system and launch further network attacks. The larger the number of Honeypots deployed, the more likely that one can effectively gather information about network attacks or probes. For example, most of the TCP connection requests are generated and sent to randomly selected IP addresses. One can identify that the connection request is for malicious traffic only after a TCP handshake is over and payload containing malicious contents is received. The probability of this happening will be more as the number of Honeypots increases.

A Honeynet normally employs a Honeywall [11], which acts as a firewall to protect the outside world from attacks emanating from within the Honeynet. In order to protect non-Honeypot systems with attacks originating from a compromised Honeypot, a Honeywall could be setup with several data control and data capture features. A Honeywall could limit the number of outbound connections allowed per hour. It could limit the bandwidth available to the attacker's traffic. A Honeywall can also on-the-fly modify malicious data packets that specifically target vulnerabilities on other systems and make them benign.

A Honeywall can capture and monitor all data traffic entering, leaving, or inside the Honeynet. The captured data can be used to analyze the steps an attacker took to compromise a Honeypot and how a compromised Honeypot is being used further. In order to prevent a Honeywall from intercepting the communications, an attacker may sometimes install encryption software to encrypt all communications between the attacker's machine and the Honeypot. In such situations, a Honeywall may not be capable of decrypting the communications. To address this problem, a logging software called Sebek [10] has been recently developed. Sebek is a software tool running on each Honeypot and is part of the machine's operating system. Sebek merely intercepts the data after the attacker's encryption software decrypts it. The decrypted information is sent to a remote server for further analysis.

### 6.1 Serial Vs Parallel Architecture

A Honeynet could be setup based on either a serial architecture (refer Figure 11) or a parallel architecture (refer Figure 12) [24]. With a serial architecture, the Honeywall filters all traffic forwarded to the firewall of the production system (the network of machines that provide useful services) and all traffic in and out of the production system goes through the Honeynet. If a Honeynet is compromised, the firewall of the production system is alerted. Even if the attack is in progress in the Honeynet, the firewall will make sure the attack does not expand into the production system. When sufficient evidence is collected in the Honeynet, one can stop running the Honeynet and start analyzing the malicious traffic directed towards it. The advantage of a Honeynet with a serial architecture is that it protects the production system from direct malicious attacks. The disadvantage would be the delay suffered for every incoming and outgoing packet, including the genuine benign ones, at the Honeynet/ firewall.

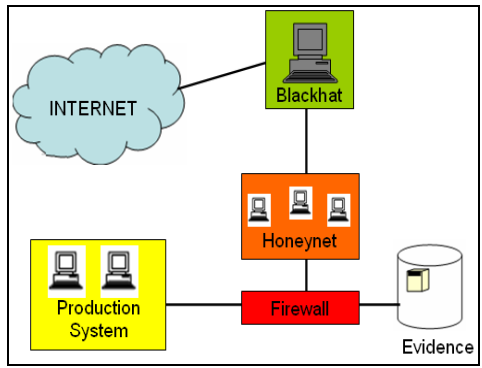
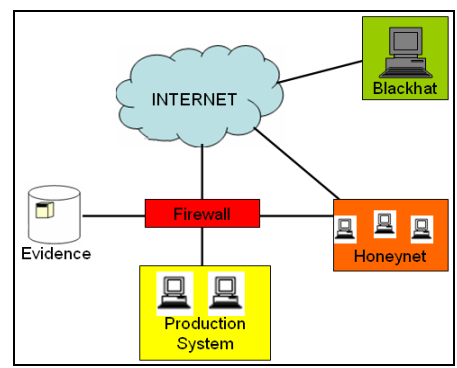

**Figure 11.** Serial Architecture        **Figure 12.** Parallel Architecture

With a parallel architecture, both the Honeynet and the production system are exposed directly to the Internet so that the delay incurred for every packet at the Honeynet/ firewall could be avoided. On the other hand, a blackhat could directly attack the production system and hence the firewall at the production system should be configured with stringent packet filtering policies to reduce the chances of untoward attacks. If the Honeynet gets some traffic, it analyzes that traffic and updates the firewall depending on the nature of the traffic received. The firewall will then become more stringent in protecting the production system.

### 6.2 Virtual Honeypots

A physical Honeypot is a real machine with its own IP address. A virtual Honeypot [25] is a simulated machine that can be modelled to behave as required. Multiple virtual Honeypots, each running the same or different operating systems, can be simulated on a single system. Each of these virtual Honeypots can be configured with specific services that will be exploited by the attackers to gain access to a Honeypot. As attackers attempt to contact machines with randomly chosen IP addresses, the more the number of virtual Honeypots deployed with the same kind of services, the more it helps to capture the connection requests and payload traffic generated by the attackers.

Honeyd [25] is a framework for virtual Honeypots to simulate computer systems at the network level. It creates virtual networks consisting of arbitrary routing topologies with configurable link characteristics like delay, bandwidth, packet loss and etc. To deceive TCP/IP stack fingerprinting tools like Xprobe [26] or Nmap [27] and to convince the adversaries that a virtual Honeypot is running a given operating system, Honeyd simulates the TCP/IP stack of the target operating system. A virtual Honeypot responds to network requests according to the services that are configured in it.

## 7. CONCLUSIONS AND FUTURE WORK

The overall contribution of this paper is an exhaustive survey of the several tools and techniques available to conduct network forensics. All the tools surveyed in this paper are free to use, at least available for trials. The paper explored in detail the different IP traceback mechanisms. Simulations were run to find out the convergence time for attack paths with different lengths and attack routers with different probabilities of marking. Finally, the paper described the Honeynet Architecture and the use of Honeypots, both and physical and virtual, in detecting malicious attack traffic and protecting the production systems.

In general, the security and forensic personnel need to keep up pace with the latest attack tools and techniques adopted by the attackers. With freely available tools, one can enforce the security mechanisms and analyze attack traffic only to a certain extent. To detect all kinds of attacks and conduct a comprehensive forensic analysis, one would have to deploy and analyze the effectiveness of commercial tools. This is the plan for future research. Future work would also involve exploring the tools and techniques available for wireless network forensics.

**Authors**

Dr. Natarajan Meghanathan is an Assistant Professor of Computer Science at Jackson State University. He graduated with MS and PhD degrees in Computer Science from Auburn University and The University of Texas at Dallas in August 2002 and May 2005 respectively. His research interests are: Ad hoc Networks, Network Security and Graph Theory. He has received grants from NSF and ARL.

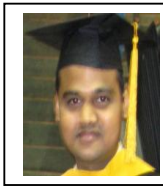

Mr. Sumanth Reddy Allam was a graduate student in Computer Science at Jackson State University. He graduated with MS degree in May 2008.

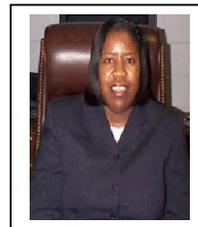

Dr. Loretta A. Moore is the Professor and Chair of the Department of Computer Science at Jackson State University, Jackson, MS. She graduated with MS and PhD Degrees in Computer Science from the Illinois Institute of Technology in 1986 and 1991 respectively. Her research interests are: Intelligent Systems, Software Systems Design, Data and Information Management, Computer Security and Forensics. She has received grants from NSF, ARL and several other federal/state agencies.